\documentstyle{article}

\newcommand{\beq}{\begin{equation}}
\newcommand{\eeq}{\end{equation}}
\newcommand{\bea}{\begin{eqnarray}}
\newcommand{\eea}{\end{eqnarray}}
\newcommand{\bi}{\bibitem}
\newcommand{\p}{\partial}

\begin{document}

\begin{center}
              {\LARGE
                     On Spacetimes Admitting Shear-free, \\
                     Irrotational, Geodesic Timelike Congruences}
\end{center}

\begin{center}{\Large
                     Alan A. Coley and Des J. McManus}
\end{center}

% \vspace{2cm}

\begin{center}{\sl
                 Department of Mathematics, Statistics and Computing Science,\\
                 Dalhousie University,
                 Halifax, NS, Canada B3H 3J5
              }
\end{center}

\vspace{.5cm}

%% FOLLOWING LINE CANNOT BE BROKEN BEFORE 80 CHAR
%%%%%%%%%%%%%%%%%%%%%%%%%%%%%%%%%%%%%%%%%%%%%%%%%%%%%%%%%%%%%%%%%%%%%%%%%%%%%%%%
%                         ABSTRACT
%%%
%
%%%
%% FOLLOWING LINE CANNOT BE BROKEN BEFORE 80 CHAR
%%%%%%%%%%%%%%%%%%%%%%%%%%%%%%%%%%%%%%%%%%%%%%%%%%%%%%%%%%%%%%%%%%%%%%%%%%%%%%%%

\begin{center}{\large {\bf Abstract}} \end{center}

\noindent
A comprehensive analysis of general relativistic
spacetimes which admit a shear-free, irrotational
and geodesic timelike congruence is presented.  The equations governing the
models for a general energy-momentum tensor are written down.  Coordinates in
which the metric of such spacetimes takes on a simplified form are established.
The general subcases of `zero anisotropic stress', `zero heat flux vector'
and  `two component fluids' are investigated.  In particular,
perfect fluid Friedmann-Robertson-Walker models
and spatially
homogeneous models are discussed.  Models with a variety of physically
relevant energy-momentum tensors are  considered.
Anisotropic fluid models and  viscous fluid models with heat
conduction are examined. Also, models with a perfect fluid plus
a magnetic field or with pure radiation, and models with two non-collinear
perfect fluids (satisfying a variety of physical conditions) are investigated.
In particular,
models with a (single) perfect fluid which is tilting with respect to the
shear-free, vorticity-free and acceleration-free timelike congruence are
discussed.

\vspace{1cm}

\begin{center} PACS: 04.20.Jb; 98.80.Hw \end{center}

%\noindent Short Title: Shear-free, Irrotational, Geodesic Timelike Congruences

 \noindent To appear in {\it Classical and Quantum Gravity} 1994
\pagebreak

%%%%%%%%%%%%%%%%%%%%%%%%%%%%%%%%%%%%%%%%%%%%%%%%%%%%%%%%%%%%%%%%%%%%%%%%%%%%%%%
%                                                                             %
                          \section{Introduction}                              %
%                                                                             %
%%%%%%%%%%%%%%%%%%%%%%%%%%%%%%%%%%%%%%%%%%%%%%%%%%%%%%%%%%%%%%%%%%%%%%%%%%%%%%%

In this paper, we shall consider general relativistic
spacetimes which admit a shear-free,
irrotational and geodesic (SIG) timelike congruence.  It is well known
\cite{ELLIS,COLLINS}  that if the stress-energy tensor is a perfect fluid
whose flow lines form a SIG
timelike congruence and whose density and pressure, as measured by a comoving
observer, satisfy a
barotropic equation of state, $p = p(\mu)$ (with $\mu + p \neq 0$), then
the spacetime must be a Friedmann-Robertson-Walker (FRW) model.
FRW spacetimes  admit a six-parameter isometry group, $G_6$,
whose orbits form spacelike hypersurfaces of constant curvature;
coordinates can be
chosen in such spacetimes so that the metric may be written as:
\beq
    ds^2 \;=\; - dt^2 \,+\, H^2(t) \, \frac{ dr^2 \,+\, r^2 \,
		 (d\theta^2 \,+\, \sin^2\theta \, d\phi^2)}
		 {(1 \,+\, \frac{\kappa}{4}\, r^2)^2}  \;\;\;,
\eeq
where the constant $\kappa$ has been normalised to take the
values $\pm 1$ and $0$.
It is also known \cite{ELLIS,COLEY&TUPPER_A} that if
an equation of state $p = p(\mu)$ is not assumed a priori then the
resulting spacetime is still FRW.
However, in the case of  more general stress-energy tensors, there do exist
solutions of the Einstein field equations  that are
not necessarily FRW models but where the spacetime admits  a SIG
timelike congruence.

It is our aim here to explore the richness of spacetimes admitting a SIG
timelike congruence.
For illustration, we begin by presenting an example of a non-FRW spacetime
that contains a SIG timelike congruence.
Consider the following  Bianchi  VI${}_0$ spacetime with
``equal scale
factors''
(spatially homogeneous spacetimes will be
studied in detail later):
\beq
    ds^2 \;=\; -dt^2 \,+\, X^2(t) (dx^2 \,+\, e^{-2x} \,dy^2 \,+\, e^{2x}\,
dz^2) \;\;\;.
    \label{1.1}
\eeq
Taking the source to be a fluid with four-velocity $u^a = \delta^a_t$, we
observe that the fluid flow lines form a SIG timelike congruence.
However, the source \underline{cannot} be a perfect fluid
since the Bianchi VI${}_0$
models do not contain FRW models as special cases.
Indeed,  this model can be interpreted as an anisotropic fluid with
\bea
    \mu &=& \frac{3\dot{X}^2}{X^2} - \frac{1}{X^2}\;\;\;, \\
    p_{\parallel} &=& -\left[ \frac{2 \ddot{X}}{X}  \,+\,
                      \frac{\dot{X}^2}{X^2} \,+\, \frac{1}{X^2}\right]\;\;\;,
\\
    p_\perp &=& -\left[ \frac{2 \ddot{X}}{X}  \,+\,
                \frac{\dot{X}^2}{X^2} - \frac{1}{X^2}\right]
		\neq p_\parallel \;\;\;,
    \label{1.1a}
\eea
where $\mu$ is the energy density, and $p_\parallel$ and $p_\perp$ are,
respectively, the pressures parallel to and perpendicular to
$n_a = X \delta_a^x$.
Hence, there exists an anisotropic fluid solution whose flow lines are SIG.
This example refutes the belief that
spatially homogeneous spacetimes admitting a SIG timelike congruence
must
necessarily be FRW models.

We are primarily interested in cosmological models.
In such models (and, in fact,
in all models) other physical conditions must be satisfied.  The precise
conditions
depend on the particular energy-momentum tensor considered; for
example, for viscous fluid models with heat conduction a suitable set of
thermodynamical laws must also be satisfied.  All models should satisfy the
(various) energy conditions \cite{HAWKING&ELLIS}.  We note that
there exist models in the above example in which the weak, dominant and strong
energy conditions are all satisfied. Consequently, physical constraints
(or, at least, the energy conditions alone) will not force the models to be
FRW.

The remainder of the paper is organised as follows.  In the next section the
governing equations are displayed.  For a spacetime in which there exists
a shear-free, irrotational, geodesic timelike congruence, coordinates can be
chosen so that the metric takes on a simplified form [cf. equations
(\ref{2.19}) and (\ref{2.20})]; the governing equations are then given
in these
coordinates.  In the following two sections general spacetimes
are investigated with zero anisotropic stress and zero heat flux. At the end of
section three, we show that in the
special case of a perfect fluid whose flow forms a SIG timelike congruence
that the density and pressure automatically satisfy a barotropic equation
of state.  In section 5, anisotropic fluid
models are considered.
In section 6, the two special physical
subcases
of a viscous fluid with heat conduction (satisfying a set of phenomenological
equations) and a perfect fluid plus a magnetic field are
investigated.  In section 7, two component fluid models are considered, and
particular attention is paid to the special physical subcases of a perfect
fluid with pure radiation and two non-collinear perfect fluids (satisfying a
variety of physical conditions such as, for example, separate energy
conservation and linear equations of state).  Finally, a single `tilting'
perfect fluid is considered.  We conclude with a discussion.

\setcounter{equation}{0} % Reset the equation counter

%%%%%%%%%%%%%%%%%%%%%%%%%%%%%%%%%%%%%%%%%%%%%%%%%%%%%%%%%%%%%%%%%%%%%%%%%%%%%%%
%
                  \section{General Equations}
%
%%%%%%%%%%%%%%%%%%%%%%%%%%%%%%%%%%%%%%%%%%%%%%%%%%%%%%%%%%%%%%%%%%%%%%%%%%%%%%%

\noindent
Einstein's field equations are \footnote{We shall follow the notation
and conventions in Ellis \cite{ELLIS}; in particular,
all kinematical quantities
are defined therein.  Also, Latin indices range
from $0$ to $3$ and Greek indices from $1$ to $3$, and
subscripts indicate differentiation with
respect to the
relevant spacetime coordinates.}
\beq
    G_{ab} \;=\; T_{ab} \label{2.1}
\eeq
where the stress-energy tensor can formally
be decomposed with respect to a timelike vector
field $u^a$ according to
\beq
    T_{ab} \;=\; \mu \,u_a\, u_b \,+\, p \, h_{ab} \,+\, q_a\, u_b
    \,+\, q_b \,u_a \,+\, \pi_{ab} \;\;\;, \label{2.2}
\eeq
where
\beq
    q_a u^a \;=\; 0, \quad \pi^a{}_a = 0, \quad \pi_{ab} u^b = 0
    \label{2.3} \;\;\;,
\eeq
and the projection tensor is defined by
$h_{ab} = g_{ab} \,+\, u_a u_b$.  In this formal
decomposition $\mu, \,p, \,q_a$ and $\pi_{ab}$
are given by
\bea
    \mu &=&  T_{ab} \,u^a\, u^b\;\;\;, \label{2.4}\\
      p &=& \frac{1}{3}\, h^{ab}\, T_{ab}\;\;\;, \label{2.5}\\
    q_a &=&  -h_a{}^c \, T_{cd} \, u^d\;\;\;, \label{2.6} \\
    \pi_{ab} &=& h_a{}^c\, h_b{}^d\, T_{cd} - \frac{1}{3}
                 \,(h^{cd}\, T_{cd}) \,h_{ab}\;\;\;. \label{2.7}
\eea
For a fluid with four-velocity $u^a$, these
quantities denote the energy density, pressure, heat conduction and anisotropic
stress, respectively, as measured
by an observer moving with the
fluid.

In this paper, we consider spacetimes in which
the shear, vorticity and acceleration (see \cite{ELLIS} for definitions)
of $u_a$ are all zero, whence the covariant derivative of $u_a$ can be
written as
\beq
    u_{a;b} \;=\; \frac{1}{3}\, \theta \,h_{ab} \label{2.8} \;\;\;,
\eeq
where $\theta \equiv u^a{}_{;a}$ is the expansion.  The relevant equations in
the case of zero shear, zero vorticity and zero acceleration are \cite{ELLIS}:

\noindent
the conservation equations
\beq
    \dot{\mu} \,+\, (\mu \,+\, p) \,\theta \,+\, q^a{}_{;a} \;=\; 0 \;\;\;,
    \label{2.9}
\eeq
\beq
    h_a{}^b \,(p_{,b} \,+\, \pi_b{}^c{}_{;c} \,+\,
    \dot{q}_b \,+\, \frac{4}{3} \,
    \theta \, q_b) \;=\; 0\;\;\;; \label{2.10}
\eeq
the Raychaudhuri equation
\beq
    \dot{\theta} \,+\, \frac{1}{3} \,\theta^2 \,+\, \frac{1}{2} \,
    (\mu \,+\, 3p) \;=\; 0
    \label{2.11} \;\;\;;
\eeq
the propagation equation
\beq
    \pi_{ab} \;=\; 2\, E_{ab}\;\;\;; \label{2.12}
\eeq
the constraint equations
\bea
   \frac{2}{3} \,\theta_{,b} \,h^b{}_a &=& q_a\;\;\;, \label{2.13}\\
    H_{ab} &=& 0\;\;\;; \label{2.14}
\eea
and the Bianchi identities (using the above)
\bea
    \pi_{tb}{}^{;b} &=& \frac{1}{3} \,h_t{}^b\, \mu_{,b} \,-\, \frac{1}{3}\,
    \theta \,q_t \;\;\;, \label{2.15} \\
    \dot{\pi}_{tm} \,+\, \frac{2}{3} \,\theta \, \pi_{tm} &=& -\frac{1}{2}\,
    h_t{}^a \, h_m{}^c\,  q_{(a;c)} \,+\, \frac{1}{6} \,q^a{}_{;a} \, h_{mt}
    \;\;\;. \label{2.16}
\eea
The tensors $E_{ab}$ and $H_{ab}$ are
the electric and the magnetic parts of the
Weyl tensor, $C^a{}_{bcd}$, and are given by
\bea
    E_{ab} &=& C_{acbd} \,  u^c\, u^d\;\;\;, \label{2.17} \\
    H_{ab} &=&  \frac{1}{2} \,C_{acst}\, u^c \, \eta^{st}{}_{bd}\, u^d
    \;\;\;. \label{2.18}
\eea
In the above, an overdot denotes differentiation along the fluid flow lines;
for example, $\dot{\theta} \equiv \theta_{,a}\, u^a.$

The conditions of zero acceleration and zero
vorticity imply that it is possible to choose a comoving coordinate system
such that
$u^a = \delta^a_t$  and such that the metric may be written as
\beq
    ds^2 \;=\; -dt^2 \,+\, g_{\alpha \beta} (t, x^\gamma)
    \, dx^\alpha \,dx^\beta
    \;\;\;. \label{2.19}
\eeq
The shear-free condition then implies that
\beq
    g_{\alpha \beta}  \;=\; H^2 (t, x^\gamma) \, h_{\alpha \beta} (x^\gamma)
    \;\;\;, \label{2.20}
\eeq
and hence the expansion is $\theta = 3 \dot{H}/H$.  We note that coordinates
can also be chosen so that $h_{\alpha \beta}$ is diagonal.

Also, the Gauss-Codazzi equations imply that the Ricci tensor of the
three-space of constant $t$ is
\beq
    R^*_{\alpha \beta} \;=\; \pi_{\alpha \beta} \,+\, \frac{1}{3} \,
    g_{\alpha \beta} \, (2 \mu \,-\, \frac{2}{3} \,\theta^2) \label{2.21}
\eeq
which yields
\beq
    R^* \;=\; 2 \mu \,-\, \frac{2}{3}\, \theta^2\;\;\;. \label{2.22}
\eeq

In the coordinates $(t, x^\alpha)$ of (\ref{2.19}),
$u^a = (1, 0, 0, 0)\;,\; q_0 =0 \;,\; \pi_{0a} = 0 \;,\;
\dot{\theta} = \p_t \theta$, and $h_a{}^b \psi_{,b} =0$ implies
$\psi_{, \alpha} = 0$.  We shall
adopt the coordinates $(t, x^\alpha)$ for the remainder of the
paper.  The energy density, the pressure, the heat flux and the anisotropic
stress tensor are then given by
\bea
    \mu &=& \frac{1}{2}\,  R^* \,+\, 3 \left(\frac{\dot{H}}{H} \right)^2
    \;\;\;, \label{2.23} \\
    p &=& -\frac{1}{6} \, R^* \,-\, \frac{2 \ddot{H}}{H} \,-\,
    \left(\frac{\dot{H}}{H}\right)^2 \;\;\;, \label{2.24} \\
    q_\alpha &=&  2 \, (\dot{H}/H)_{,\alpha} \;=\; 2\,
                  \p_t [\p_\alpha ({\rm ln} H)] \;=\; 2\, \p_\alpha
                  [\p_t({\rm ln}  H)] \;\;\;, \label{2.25} \\
    \pi_{\alpha \beta} &=&  R^*_{\alpha\beta} \,-\, \frac{1}{3} \,
			    g_{\alpha\beta}\,  R^* \;\;\;. \label{2.26}
\eea
Equations (\ref{2.23}) and (\ref{2.26}) follows directly from (\ref{2.21})
and (\ref{2.22}).  Equation (\ref{2.24}) follows by inserting (\ref{2.23})
into the
Raychaudhuri equation (\ref{2.11}).  The conservation equations (\ref{2.9})
and
(\ref{2.10}) become
\bea
    \dot{\mu} \,+\, \frac{3 \dot{H}}{H} \, (\mu \,+\, p) \,+\,
    g^{\alpha \beta} \, q_{\alpha;\beta} &=&  0\;\;\;, \label{2.27} \\
    \p_t(q_\alpha) \,+\, \frac{3 \dot{H}}{H} \, q_\alpha \,+\,
    p_{,\alpha} \,+\,
    g^{\beta\gamma}\,  \pi_{\alpha\beta;\gamma} &=& 0 \;\;\;. \label{2.28}
\eea
Now, the three-metric $g_{\alpha\beta}$ is conformal
to the metric $h_{\alpha\beta} \;,\; g_{\alpha\beta} = H^2\, h_{\alpha\beta}$,
and
thus the Ricci tensor $R^*_{\alpha\beta}$ is related to
the Ricci tensor ${}^3R_{\alpha\beta}$, of the metric $h_{\alpha\beta}$
\cite{HAWKING&ELLIS} by
\bea
    R^*_{\alpha\beta} &=& {}^3R_{\alpha\beta} \,-\, H^{-1} \nabla_\alpha
                          \nabla_\beta H \,-\, H^{-1}
                          (\nabla^2 H)\, h_{\alpha\beta} \,+\, 2H^{-2}
			  \nabla_\alpha H \, \nabla_\beta H \label{2.29} \\
                      &=& {}^3R_{\alpha\beta} \,-\, \nabla_\alpha \nabla_\beta
		      ({\rm ln} H) \,+\, \nabla_\alpha({\rm ln} H)  \nabla_\beta
		      ({\rm ln} H) \nonumber \\
                      & &  - [\nabla^2 ({\rm ln} H) \,+\,  \underline{\nabla}
		           ({\rm ln} H) \cdot
                           \underline{\nabla} ({\rm ln} H)]\, h_{\alpha\beta}
			   \;\;\;, \label{2.30}
\eea
where $\nabla^2 H \equiv h^{\alpha\beta}
\nabla_\alpha \nabla_\beta H \;,\;
\underline{\nabla} H \cdot  \underline{\nabla} H \equiv h^{\alpha\beta}
\nabla_\alpha H \nabla_\beta H \;,\;
h^{\alpha\beta}\,  h_{\beta\gamma} = \delta^\alpha_\gamma$,
and $\nabla_\alpha$ is the
covariant derivative with respect to the metric $h_{\alpha\beta}$.
Thus, re-expressing (\ref{2.23})--(\ref{2.28}) in terms of the three-metric
$h_{\alpha\beta}$, we find
\bea
    \mu &=& \frac{1}{2} \,H^{-2} \,({}^3R) \,-\,2 H^{-3} \,\nabla^2 H \,+\,
            H^{-4}
            \underline{\nabla} H \cdot  \underline{\nabla} H  \,+\, 3H^{-2}
            \dot{H}^2
             \;\;\;, \label{2.31} \\
      p &=& -2 H^{-1} \ddot{H} \,- \, \frac{1}{3}\, \mu \;\;\;, \label{2.32} \\
     q_\alpha &=& 2 \p_t [\nabla_\alpha ({\rm ln} H)] \;\;\;, \label{2.33} \\
     \pi_{\alpha\beta} &=& {}^3R_{\alpha\beta} \,-\, \frac{1}{3} \,
                           h_{\alpha\beta }\, {}^3R \,-\,
                           \nabla_\alpha \nabla_\beta ({\rm ln} H)
                           \,+\, \nabla_\alpha ({\rm ln} H)\,
                           \nabla_\beta({\rm ln} H)   \nonumber \\
      & &  \,+\,\, \frac{1}{3}\, [ \nabla^2 ({\rm ln} H) \,-\,
            \underline{\nabla}
           ({\rm ln} H) \cdot \underline{\nabla} ({\rm ln} H)] \, h_{\alpha
\beta}
           \;\;\;, \label{2.34}
\eea
and
\bea
    \dot{\mu} \,+\, 3 H^{-1} \dot{H} \, (\mu \,+\, p)  \,+\,  H^{-1}
    \nabla^\alpha q_\alpha &=& 0  \;\;\;, \label{2.35} \\
    \p_t q_\alpha \,+\, 3 H^{-1}\, \dot{H} \, q_\alpha \,+\,
    \nabla_\alpha p
    \,+\,  H^{-2} \,\nabla^\beta \pi_{\alpha\beta} &=& 0 \;\;\;, \label{2.36}
\eea
where $ \nabla^\alpha \equiv h^{\alpha\beta} \nabla_\beta$.

Finally, the Bianchi identities (\ref{2.15}) and (\ref{2.16}) reduce to
\bea
    & &H^{-1} \, \nabla^\beta \pi_{\alpha\beta}  \,+\, H^{-3}\,
    (\nabla^\beta H)
    \pi_{\alpha \beta} \;=\;  \frac{1}{3}\, \nabla_\alpha \mu \,-\, H^{-1}
                            \dot{H} \, q_\alpha \;\;\;, \label{2.37} \\
    & & 2 \p _t \pi_{\alpha \beta} \;=\;  - \nabla_\alpha q_\beta \,+\, 2
                                   q_{(\alpha}
                                   \nabla_{\beta)} ({\rm ln} H) \,+\,
                                   \frac{1}{3} \,H^2 \,\nabla^\gamma \,(H^{-2}
                                   q_\gamma) \, h_{\alpha\beta}
                                   \;\;\;. \label{2.38}
\eea
Equation (\ref{2.37}) is equivalent to the contracted Bianchi identity
$\nabla^\alpha ({}^3R_{\alpha\beta}) = \frac{1}{2} \nabla_\beta ({}^3R)$, and
(\ref{2.38}) simply expresses the result that
$\p_t({}^3R_{\alpha\beta} - \frac{1}{3} h_{\alpha \beta} \,{}^3R) = 0$.

This completes our general analysis and any further progress can only
be made on a case by case basis.

\setcounter{equation}{0} % Reset the equation counter

%%%%%%%%%%%%%%%%%%%%%%%%%%%%%%%%%%%%%%%%%%%%%%%%%%%%%%%%%%%%%%%%%%%%%%%%%%%%%%%
%                                                                             %
                      \section{Zero Anisotropic Stress}                       %
%                                                                             %
%%%%%%%%%%%%%%%%%%%%%%%%%%%%%%%%%%%%%%%%%%%%%%%%%%%%%%%%%%%%%%%%%%%%%%%%%%%%%%%

In this section, we take the anisotropic stress tensor to be zero,
$\pi_{ab} = 0$.  The equations (\ref{2.9})--(\ref{2.16}) and
(\ref{2.21})--(\ref{2.22}) then become
\bea
    &  & \dot{\mu} \,+\, (\mu \,+\, p) \theta \,+\, q^\alpha{}_{;\alpha} \;=\;
0
         \;\;\;, \label{3.1} \\
    &  & p_{,\alpha}  \,+\, \dot{q_\alpha} \,+\, \frac{4}{3} \,\theta \,
         q_\alpha \;=\; 0 \;\;\;, \label{3.2} \\
    &  & \dot{\theta} \,+\, \frac{1}{3}\, \theta^2 \,+\, \frac{1}{2}\,
         (\mu \,+\, 3p) \;=\; 0 \;\;\;, \label{3.3} \\
    & & E_{ab} \;=\; H_{ab} \;=\; 0 \;\;\;, \label{3.4} \\
    & & \frac{2}{3} \, \theta_{,\alpha} \;=\; q_\alpha \;\;\;, \label{3.5}\\
    & & \mu_{,\alpha} \;=\; \theta\, q_\alpha \;\;\;, \label{3.6} \\
    & & \frac{1}{2} \, (q_{\alpha;\beta} \,+\, q_{\beta;\alpha}) \;=\;
        \frac{1}{3} \, q^\gamma{}_{;\gamma}\,  g_{\alpha\beta}
        \;\;\;, \label{3.7} \\
    & & R^*_{\alpha\beta}  \;=\; \frac{1}{3} \, g_{\alpha\beta}
        \left(2 \mu \,-\, \frac{2}{3} \,\theta^2\right) \;\;\;, \label{3.8}\\
    & & R^*  \;=\; 2 \mu \,-\, \frac{2}{3} \, \theta^2 \;\;\;.\label{3.8a}
\eea
Equation (\ref{3.4}) implies that the Weyl tensor is zero and thus the
spacetime is conformally flat.  Equation (\ref{3.6}) can be integrated
with the aid of (\ref{3.5}) to obtain
\beq
    \mu = \frac{1}{3} \theta^2 \,+\, \frac{1}{2} f(t) \;\;\;, \label{3.9}
\eeq
where $f$ is an arbitrary function of integration.  Comparing (\ref{3.9})
with (\ref{3.8a}), we find
\beq
    R^* \;=\; f(t) \;\;\;. \label{3.10}
\eeq
Therefore, each three-space ($t =$ const) is a space of constant
curvature and thus, we may choose coordinates so that
the metric has the form \cite{BANERJEE,EISENHART,BONA&COLL}
\bea
    ds^2 &=&  -dt^2 \,+\, \Omega^{-2}\, (dx^2 \,+\, dy^2 \,+\, dz^2) \;\;\;,
               \label{3.11} \\
   \Omega(t, x^\alpha) &=&  a(t) \, (x^2 \,+\, y^2 \,+\, z^2) \,+\,
                            b_\alpha(t) \, x^\alpha \,+\, c(t) \;\;\;,
                            \label{3.12}
\eea
where $a, b_\alpha$ and $c$ are arbitrary
functions of $t$.  The Ricci scalar is then
\bea
    R^* &=& 6 \,(4 a c \,-\, \delta^{\alpha\beta}\, b_\alpha \,b_\beta)
              \;=\; R^*(t) \;\;\;, \label{3.13} \\
     &=& 4 \Omega \nabla^2 \Omega \,-\, 6 \underline{\nabla} \Omega \cdot
         \underline{\nabla} \Omega \label{3.14}
\eea
[$\underline{\nabla}$ is the
covariant derivative with respect to the
three-metric $h_{\alpha \beta} ( = \delta_{\alpha \beta})$].  The
expansion is now given as
\beq
    \theta \;=\; - \frac{3 \dot{\Omega}}{\Omega} \;\;\;. \label{3.15}
\eeq
Equation (\ref{3.7}) reduces to
$\dot{\Omega}_{,\alpha\beta} = \frac{1}{3} (\nabla^2 \dot{\Omega})
   \delta_{\alpha \beta}$
which is trivially satisfied by $\Omega$
as given
in (\ref{3.12}).  Also (\ref{3.2}) follows directly from
(\ref{3.8a}) and the spatial gradient of (\ref{3.3}).

The density, $\mu$, the pressure, $p$, and
the heat flux, $q_\alpha$, can all be expressed
in terms of $\Omega$ and its derivatives:
\bea
    \mu &=& 2 \Omega \,\nabla^2 \Omega \,-\, 3 \underline{\nabla} \Omega \cdot
            \underline{\nabla} \Omega \,+\, 3 \left(
\frac{\dot{\Omega}}{\Omega}
            \right)^2 \;\;\;, \label{3.16} \\
      p &= & 2\left( \frac{\dot{\Omega}}{\Omega} \right)^\cdot \,-\,
             3 \left( \frac{\dot{\Omega}}{\Omega} \right)^2 \,- \,
             \frac{2}{3} \, \Omega \, \nabla^2 \Omega \,+\, \underline{\nabla}
             \Omega \cdot \underline{\nabla} \Omega \;\;\;, \label{3.17} \\
     q_\alpha &=& -2 \nabla_\alpha \left( \frac{\dot{\Omega}}{\Omega} \right)
                  \;\;\;. \label{3.18}
\eea
The conservation equation (\ref{3.1}) becomes
\beq
    \dot{\mu} \,-\, 3\,(\mu \,+\, p) \frac{\dot{\Omega}}{\Omega} \,-\,
    \frac{1}{2} \, \Omega^2 \, (\Omega^{-2}\, R^*) \;=\; 0
    \;\;\;, \label{3.19}
\eeq
which is automatically satisfied by $\mu$ and $p$ as given in (\ref{3.16})
and (\ref{3.17}).

The spherically symmetric solutions are found by setting $b_\alpha (t)$
equal to zero in (\ref{3.12}) and were originally discussed by Maiti
\cite{MAITI}.
The metric can then be written as
\beq
    ds^2 \;=\; -dt^2 \,+\, \frac{dr^2 \,+\, r^2 d\theta^2 \,+\, r^2
                \sin^2\theta \, d\phi^2}{[F(t) \, r^2 \,+\, G(t)]^2} \;\;\;,
		\label{1.2}
\eeq
where $F$ and $G$ are arbitrary functions.  The model can then
be interpreted as a perfect fluid with heat conduction, where
the fluid four-velocity is $u^a = \delta^a_t$. The density, pressure,
and heat flux are now given by
\bea
    \mu &=& 3 (\dot{F} r^2 \,+\,\dot{G})^2 (F r^2 \,+\, G)^{-2} \,+\, 12 FG
    \label{1.4} \;\;\;, \\
    p &=& 2 (\ddot{F} r^2 \,+\, \ddot{G})(F r^2 \,+\, G)^{-1} -
	  5 (\dot{F} r^2 \,+\, \dot{G})^2
            (Fr^2 \,+\, G)^{-2} - 4FG \label{1.5} \;\;\;, \\
    q_1 &=& -4r [\dot{F} G - \dot{G} F] (F r^2 \,+\, G)^{-2}\;\;\;, \, q_2 =
q_3 = 0
            \;\;\;, \label{1.6}
\eea
where $q_1$ is the heat conduction in the radial direction.
We note that the model is conformally flat.  Cosmological models of
this type have been studied by Kolassis {\it et al.} \cite{KOLASSIS}
and Banerjee {\it et al.} \cite{BANERJEE}.

%%%%%%%%%%%%%%%%%%%%%%%%%%%%%%%%%%%%%%%%%%%%%%%%%%%%%%%%%%%%%%%%%%%%%%%%%%%%%%%
                                                                             %
\noindent{\bf Perfect Fluid}                              %

If both the anisotropic stress tensor and the
heat flux vector are zero $(\pi_{ab} =q_a =0)$ then the stress-energy
tensor (\ref{2.2}) is that of a perfect fluid and   the metric is the
FRW metric \cite{ELLIS}.  Thus if the stress-energy tensor is a perfect fluid,
whose flow lines form a SIG timelike congruence, then the solutions of
the Einstein field equations must be the FRW models.
Furthermore, equations (\ref{3.2}) and (\ref{3.6}) imply that both $\mu$ and
$p$ are functions of the single variable $t$, [that is,
$\mu = \mu(t) \,p = p(t)$].
Hence $\mu$ and $p$ satisfy a barotropic equation of state $p =p(\mu)$.
Therefore, $p = p(\mu)$ is a consequence of the assumptions and need not
be specified a priori.

\setcounter{equation}{0} % Reset the equation counter

%%%%%%%%%%%%%%%%%%%%%%%%%%%%%%%%%%%%%%%%%%%%%%%%%%%%%%%%%%%%%%%%%%%%%%%%%%%%%%%
%                                                                             %
                      \section{Zero Heat Flux}                                %
%                                                                             %
%%%%%%%%%%%%%%%%%%%%%%%%%%%%%%%%%%%%%%%%%%%%%%%%%%%%%%%%%%%%%%%%%%%%%%%%%%%%%%%

Here, we shall take the heat flux to be
zero, $q_a =0$.  We immediately obtain from equation (\ref{2.13}) that
$\theta = \theta(t)$, and hence we can set $H = H(t)$ without loss of
generality.  The anisotropic stress tensor is now directly related to the
Ricci tensor of
the three-metric $h_{\alpha\beta}$; equation (\ref{2.34}) implies that
\beq
    \pi_{\alpha\beta} \;=\; {}^3R_{\alpha\beta} \,-\, \frac{1}{3} \,
                         h_{\alpha\beta}\, {}^3R \;\;\;. \label{5.1}
\eeq
The energy density and the isotropic pressure, as given by (\ref{2.31}) and
(\ref{2.32}), are
\bea
    \mu &=& 3 H^{-2} \,\dot{H}^2 \,+\, \frac{1}{2} \,H^{-2}\, ({}^3R)
            \;\;\;, \label{5.2} \\
    p &=& -2 H^{-1} \,\ddot{H} \,-\, H^{-2}\, \dot{H}^2 \,-\, \frac{{}^3R}{6}
           \;\;\;. \label{5.3}
\eea
The conservation equations
\bea
    \dot{\mu} \,+\, 3 H^{-1}\, \dot{H} \,(\mu \,+\, p) &=& 0
           \;\;\;, \label{5.4} \\
    \nabla_\alpha p \,+\, H^{-2}\, \nabla^\beta \pi_{\alpha \beta} &=& 0
          \;\;\;, \label{5.5}
\eea
and the Bianchi identities
\bea
    H^{-2} \, \nabla^\beta \pi_{\alpha\beta} &=& \frac{1}{3} \,
                  \nabla_\alpha \mu \;\;\;, \label{5.6} \\
    \p_t \pi_{\alpha\beta} &=& 0 \;\;\;, \label{5.7}
\eea
are automatically satisfied.

The zero heat flux models, under consideration in this section,
can be subdivided into three distinct
classes depending on the number of distinct eigenvalues of $\pi_{\alpha
\beta}$.  If $\pi_{\alpha \beta}$ has three distinct eigenvalues then the
spacetime is
of Petrov type I \cite{TRUMPER}.  If $\pi_{\alpha\beta}$ has exactly two
distinct eigenvalues then the spacetime is of Petrov type D
\cite{TRUMPER,FERRANDO_92}.  Finally, if $\pi_{\alpha \beta}$ has only one
eigenvalue (the eigenvalue is zero since $\pi_{\alpha\beta}$ is tracefree,
and hence $\pi_{\alpha\beta}$ is identically zero) then the spacetime is
conformally flat and corresponds to a perfect fluid FRW model.

In comoving coordinates, the metric has the
following form:
\beq
    ds^2 \;=\; -dt^2 \,+\, H^2(t)\, h_{\alpha\beta}(x^\gamma) \, dx^\alpha
               dx^\beta \;\;\;. \label{5.8}
\eeq
Thus any spatial three-metric will give rise to
a shear-free, irrotational, geodesic model with zero heat flux.  In particular,
there exist examples of spatially homogeneous spacetimes of all the Bianchi
types \cite{BIANCHI}, I--IX, and the Kantowski-Sachs solution \cite{K&S}
that admit a SIG timelike congruence.  However, only Bianchi types I, V,
VII${}_0$, VII${}_h$ and IX can give rise to FRW models.  Indeed, Mimoso and
Crawford \cite{MIMOSO&CRAWFORD} recently
drew attention to the same fact: ``No anisotropic model can
simultaneously exhibit a perfect fluid matter content
and a shear-free timelike congruence \ldots But the possibility still
exists of considering an imperfect fluid for shear-free anisotropic
models'' (where presumably the authors mean a shear-free timelike
\underline{fluid} congruence, otherwise the statement is untrue --- see
section 6).

As an illustrative example, we shall consider the following Bianchi
IV spacetime with ``equal scale factors,'' since this does not
contain any FRW models as special cases:
\beq
    ds^2 \;=\; -dt^2   \,\,+\, \, H^2(t)\, \{dx^2 \,+\, e^{2x} \,(x^2\,+\,1)\,
               dy^2 \,+\, 2x\, e^{2x}\, dydz \,+\, e^{2x}\, dz^2\}
               \;\;\;. \label{5.9}
\eeq
Taking the source to be an anisotropic fluid with the four-velocity of this
fluid given by $u^a = \delta^a_0$, we observe that the
acceleration, shear, and vorticity of the fluid congruence all vanish, and
that the heat flux as measured by a comoving observer is zero.
Using (\ref{5.1}), we find that the non-zero components of the
anisotropic stress tensor are
\bea
    \pi_{xx} &=& -\frac{1}{3} \;\;\;,  \label{5.10} \\
    \pi_{yy} &=& e^{2x} \,\left(\frac{2}{3}\, x^2 \,-\,2x \,- \, \frac{1}{3}
                 \right) \;\;\;, \label{5.11} \\
    \pi_{yz} &=& e^{2x}\, \left(\frac{2}{3} \, x \,-\, 1 \right)
                 \;\;\;, \label{5.12} \\
   \pi_{zz} & =& \frac{2}{3} \, e^{2x} \;\;\;. \label{5.13}
\eea
The eigenvalue of $\pi_{\alpha\beta}$ can be found by
solving the characteristic equation
det$(\lambda h_{\alpha\beta} - \pi_{\alpha\beta}) = 0$.
There are three distinct eigenvalues; namely,
\beq
    \lambda \;=\; \frac{1}{3} \, , \;\;\;\;\;\;
    \lambda \;=\; \frac{1}{6}\, \pm \,\frac{1}{2}\sqrt{5} \;\;\;. \label{5.14}
\eeq
Thus, the spacetime is of Petrov type I.  The energy density and the
isotropic pressure are given by (\ref{5.2}) and (\ref{5.3}), respectively,
with $R = -13/2$.

We note that for each three-metric $h_{\alpha \beta}$
there exists an anisotropic stress tensor $\pi_{\alpha \beta}$, but
not all of the $\pi_{\alpha \beta}$ are related to physical relevant
matter.  Therefore, it is of interest
to study models with some extra physical conditions.  Perhaps the most
reasonable condition to assume here is an equation of state
between the energy density, $\mu$,
and the isotropic pressure, $p$.
%%%%%%%%%%%%%%%%%%%%%%%%%%%%%%%%%%%%%%%%%%%%%%%%%%%%%%%%%%%%
%\newpage

\noindent{\bf Equation of state: $p = p(\mu)$}

\noindent
Here, we consider the case where the energy density and the isotropic pressure
satisfy an equation
of state, $p = p(\mu)$, where both $p$ and $\mu$ are
differentiable.  This implies that the integrability condition
\beq
    (\nabla_\alpha  p) \dot{\mu} \;=\; (\nabla_\alpha \mu) \dot{p} \label{5.15}
\eeq
must be satisfied.  Thus, inserting (\ref{5.2}) and (\ref{5.3}) into
(\ref{5.15}) yields
\beq
    (\nabla_\alpha \mu) \, \p_t (H^{-1} \ddot{H}) \;=\; 0
         \;\;\;. \label{5.16}
\eeq
Two possible solutions exist:  either (i) $\nabla_\alpha \mu = 0$ or  (ii)
$\p_t (H^{-1} \ddot{H}) =0$.

\noindent {\bf (i)  $\nabla_\alpha \mu =0$}

\noindent
Here $\mu = \mu(t)$, and (\ref{5.2}) implies that the Ricci scalar
is constant,  ${}^3R =$const.  The isotropic pressure is now also a function
of $t$, $p = p(t)$.  The anisotropic stress-tensor, as given by (\ref{5.1}), is
now subject to the constraint $\nabla^\alpha \pi_{\alpha\beta} = 0$.
Therefore, models with zero heat flux where the energy
density and the isotropic pressure satisfy a barotropic equation of state are
characterised by a single scale factor, $H(t)$, and a spatial three-metric
whose Ricci scalar is constant (not to be confused with three-spaces of
constant  curvature).

We now make a brief comment on the consequences of the various
energy conditions for the above models.  The weak energy condition \cite{WALD}
--- $T_{ab} v^a v^b$ is non-negative for all unit timelike vectors $v^a$ ---
implies that the energy density, $\mu$, is
non-negative and hence, if ${}^3R$ is negative then (\ref{5.2}) implies that
either $\dot H >0$  or $\dot H <0$ for all time, $t$; thus, in the former case,
the cosmology will be an open model.  The strong energy condition
--- $2T_{ab} v^a v^b \,+\, T_a \,^a$ is non-negative for all unit timelike
vectors
$v^a$ --- implies that $\mu \,+\,3p \geq 0$, and thus the deceleration
parameters,
$q \equiv -\ddot H H/ \dot H^2$, is non-negative.

\newpage

\noindent{\bf (ii) $\p_t (H^{-1}\, \ddot{H}) = 0$}

\noindent
Therefore, $\ddot H = \lambda H$ where $\lambda$ is a constant.
We can integrate this equation to get
\beq
    \dot H^2 \;=\; \lambda\, H^2 \,+\, \nu\;\;\;, \label{5.17}
\eeq
where $\nu$ is a constant.  The energy density and the isotropic pressure are
then
\bea
    \mu &=& \frac{1}{2}\,  H^{-2}\, ({}^3R \,+\, 6 \nu) \,+\, 3 \lambda
            \;\;\;, \label{5.18} \\
      p &=& -\frac{1}{3}\, \mu \,-\, 2 \lambda \;\;\;. \label{5.19}
\eea
If the fluid satisfies the strong energy condition then we can set
$\lambda = -\alpha^2 \leq 0$ and
$\nu = \alpha^2 \beta^2 \geq 0$ for some constants $\alpha$ and $\beta$.
If $\alpha \ne 0$ then (\ref{5.17}) integrates to yield
\beq
    H(t) \;=\; \beta\, \sin(\alpha \,t) \label{5.20}
\eeq
which is a closed model.  If $\lambda =0$ then
\beq
    H(t) \;=\; \gamma \,t \;\;\;\;\;\; (\gamma = {\rm const}) \label{5.21}
\eeq
which is an open model.
(In both cases, $\frac{dp}{d\mu} = -\frac{1}{3}$.)

Therefore, if the heat flux of the fluid is zero and the energy density and the
isotropic pressure satisfy a barotropic equation of state then the
hypersurfaces $t =$ constant are for the most part hypersurfaces of
constant Ricci scalar with the single exception of the equation of state
$\frac{dp}{d\mu} = -\frac{1}{3}$.

%%%%%%%%%%%%%%%%%%%%%%%%%%%%%%%%%%%%%%%%%%%%%%%%%%%%%%%%%%%%%%%%%%%%%%%%%%%%%%%

\noindent{\bf Orthogonal Spatially Homogeneous Models}                  %

The function $H$ appearing in the metric (\ref{2.20}) must be independent of
the spatial
coordinates for the metric to represent a
spatially homogeneous model.  Therefore, the expansion also only
depends on $t$, $\theta = \theta(t)$,
and hence the heat flux is zero.

To make further progress, we assume that the matter associated
with these spatially homogeneous models inherits the symmetries
\cite{COLEY&TUPPER_A};
in particular, all scalars should be independent of the spatial coordinates.
Hence, both the energy density, $\mu$,  and
the isotropic pressure, $p$, are now functions of only $t$.
Equation (\ref{5.2}) then implies
that the Ricci scalar, ${}^3R$, is constant, which is case (i) for zero heat
flux models with a barotropic equation of state.  The eigenvalues of the
anisotropic stress tensor, $\pi_{\alpha \beta}$, should also be constants.
Indeed, as might be expected, all of the
Bianchi types (I--IX) and the Kantowski-Sach models admit
solutions with constant Ricci scalar such that the eigenvalues of
$\pi_{\alpha \beta}$, as given by (\ref{5.1}), are constant.

% We note that there exist solutions with a constant Ricci scalar,
% ${}^3R$, and constant eigenvalues that are not Bianchi models.

\setcounter{equation}{0} % Reset the equation counter

%%%%%%%%%%%%%%%%%%%%%%%%%%%%%%%%%%%%%%%%%%%%%%%%%%%%%%%%%%%%%%%%%%%%%%%%%%%%%%%
%                                                                             %
                      \section{Anisotropic Fluid}                             %
%                                                                             %
%%%%%%%%%%%%%%%%%%%%%%%%%%%%%%%%%%%%%%%%%%%%%%%%%%%%%%%%%%%%%%%%%%%%%%%%%%%%%%%

The stress-energy tensor for an anisotropic fluid is
\beq
    T_{ab} \;=\; \mu\, u_a\, u_b \,+\, p_{\parallel}\, n_a \, n_b \,+\,
                 p_\perp \, (u_a\, u_b \,-\, n_a \,n_b \,+\, g_{ab})
                 \;\;\;, \label{6.1}
\eeq
where $u^a$ is a unit timelike vector, $u_a u^a =-1$, and
$n^a$ is a unit spacelike vector, $n_a n^a =1$,
orthogonal to $u^a$, $u^a n_a =0$.  The scalars $p_{\parallel}$ and $p_\perp$
are the pressure parallel and perpendicular to $n^a$, respectively.
Equation (\ref{2.6}) implies that the heat flux in the case of an
anisotropic fluid is zero.  Thus, shear-free, irrotational, geodesic
anisotropic fluids are
a subcase of the models discussed in section 4.  Equations (\ref{2.5}) and
(\ref{2.7}) yield expressions for the isotropic pressure and the anisotropic
stress tensor;
\bea
    p &=& \frac{1}{3} \, (p_{\parallel} \,+\, 2 p_\perp)\;\;\;, \label{6.2}\\
    \pi_{ab} &=&  (p_\parallel  \,-\, p_\perp)\, \{ n_a \,n_b \,-\,
                  \frac{1}{3} \, (g_{ab} \,+\, u_a \, u_b)\} \;\;\;.
\label{6.3}
\eea
In terms of the comoving coordinates of (\ref{2.19}), the spacelike vector
$n_a = (0, n_\alpha)$ and
the anisotropic stress tensor is given by
\beq
    \pi_{\alpha\beta} \;=\; (p_\parallel \,-\, p_\perp) \,
                            (n_\alpha \,n_\beta \,-\, \frac{1}{3}\, H^2
                             h_{\alpha \beta})\;\;\;. \label{6.4}
\eeq
The anisotropic stress tensor (\ref{6.4}) may be decomposed as \cite{DES&ALAN}
\beq
    \pi_{\alpha\beta} \;=\; P \,(N_\alpha\, N_\beta \,-\, \frac{1}{3} \,
                            h_{\alpha \beta}) \;\;\;, \label{6.5}
\eeq
where $N_\alpha(x^\beta) \;[ = H^{-1}(t)\, n_\alpha]$ is a unit vector with
respect to the metric $h_{\alpha \beta}$ and $P$ is
independent of $t$, $P = P(x^\alpha)$.  The function $P$ is a measure of the
anisotropy of the fluid,
\beq
    P \;=\; H^2(t)\, (p_\parallel \,-\, p_\perp)\;\;\;, \label{6.6}
\eeq
and is subject to the constraint equation
\beq
    \frac{1}{6} \, \nabla_\alpha \, ({}^3R) \;=\;  \nabla^\beta
             (P \,N_\beta)\, N_\alpha \,-\, \frac{1}{3} \, \nabla_\alpha P
              \,+\,  P\, N^\beta \,\nabla_\beta N_\alpha \;\;\;. \label{6.7}
\eeq
The energy density and the isotropic pressure are given by expressions
(\ref{5.2}) and (\ref{5.3}), respectively.  The anisotropic pressures are
related to $p$ and $P$ by
\bea
    p_\perp &=&  p \,-\,  \frac{P}{3 H^2} \;\;\;, \label{6.8} \\
    p_\parallel &=& p \,+\, \frac{2P}{3 H^2} \;\;\;. \label{6.9}
\eea

The anisotropic stress tensor (\ref{6.5}) has only two distinct eigenvalues,
$\frac{2}{3}P$ and $-\frac{1}{3}P$ (double), and hence the spacetime is
of Petrov type D (if $P \ne 0$).  However, if $P$
is identically zero then (\ref{6.6}) implies that $p_\parallel = p_\perp$
and thus, the spacetime is then a perfect fluid FRW model.  Conversely, if the
metric is of the form (\ref{5.8}) and $\pi_{\alpha \beta}$ has exactly two
distinct eigenvalues then $\pi_{\alpha \beta}$ must necessarily be of the
form (\ref{6.5}) and hence the
stress-energy tensor will formally be that of an anisotropic fluid with
four-velocity $u^a = \delta^a_t$ \cite{DES&ALAN}.
Furthermore,  if $p$ and $\mu$ satisfy a barotropic equation of state, $p =
p(\mu)$, then either the Ricci scalar, ${}^3R$, is constant or
$\frac{dp}{d\mu} = -\frac{1}{3}$ (see section 4).  The case where
the Ricci scalar, ${}^3R$, is constant is the most interesting; in particular,
there exists examples of spatially homogeneous anisotropic fluid
models of Bianchi types II, III, VI and VIII, and a
Kantowski-Sachs solution \cite{DES&ALAN}.

\setcounter{equation}{0} % Reset the equation counter

%\newpage

%%%%%%%%%%%%%%%%%%%%%%%%%%%%%%%%%%%%%%%%%%%%%%%%%%%%%%%%%%%%%%%%%%%%%%%%%%%%%%%
%                                                                             %
                             \section{Special Cases}                          %
%                                                                             %
%%%%%%%%%%%%%%%%%%%%%%%%%%%%%%%%%%%%%%%%%%%%%%%%%%%%%%%%%%%%%%%%%%%%%%%%%%%%%%%

\noindent{\bf Viscous Fluid}

\noindent For a viscous fluid with heat conduction the following
phenomenological equations are satisfied:
\bea
    \pi_{ab} &=& -2 \eta \, \sigma_{ab} \;=\; 0\;\;\;, \label{8.1} \\
    q_a &=& -\chi\, h_a{}^b\, (T_{;b} \,+\, T\, \dot{u_b})
	   \;=\; - \chi \, h_a{}^b \, T_{;b} \;\;\;, \label{8.2} \\
    p &=& p_t \,-\, \zeta \, \theta \;\;\;, \label{8.3}
\eea
where $\zeta$ is the bulk  viscosity coefficient
($\eta$ is the shear viscosity coefficient), $p_t$
is the thermodynamic pressure, $\chi$ is the heat conductivity, and $T$
is the temperature.  These quantities are restricted by equations of state
of the form  $\zeta = \zeta(\mu, T)$
\linebreak
$[\eta = \eta(\mu, T)]\,,\, \chi = \chi(\mu, T)\,,\, p_t = p_t(\mu, T)$,
and the
various energy conditions.  In addition, the Gibb's relation, the baryon
conservation law and the second law of thermodynamics must be satisfied.

Since $\pi_{ab} = 0$, these models are a subcase of the models
discussed in section 3, and hence there exists coordinates in which the
metric is given by equations (\ref{3.11}) and (\ref{3.12}), and $\mu$, $p$ and
$q_\alpha$ are
given by (\ref{3.16})--(\ref{3.18}).

As an illustration, let us consider the case in which $T = T(\mu)$, whence
\linebreak$\zeta = \zeta(\mu) \,,\, \chi = \chi(\mu)$ and $p_t = p_t(\mu)$.
For example,
at the high temperatures in the early universe when the energy density
was dominated by relativistic species (that is, when the universe
was radiation
dominated), $\mu = \frac{1}{2} ga T^4 = \mu(T)$, where $g = g(T)$ is the
number of effective degrees of freedom contributing to the universe
at temperature $T$.  From equations (\ref{8.2}) and (\ref{3.6}),
we  obtain
\beq
    \frac{1}{\theta} \, \mu_{,\alpha} \;=\; -\chi \, T_{,\alpha}
	   \;=\; - \chi \, \frac{dT}{d\mu} \, \mu_{,\alpha}
	   \;\;\;, \label{8.4}
\eeq
whence
\beq
    \theta \;=\; - \left(\chi\, \frac{dT}{d\mu}\right)^{-1}
           \;=\; \theta(\mu) \;\;\;, \label{8.5}
\eeq
since $\mu_{,\alpha} \neq 0 \;\;\; (q_\alpha \neq 0$), else the spacetime
is a
perfect fluid FRW spacetime.  This implies that $f(t) =$ constant in
(\ref{3.9})
and $p = p(\mu)$ from (\ref{8.3}).  Equation (\ref{3.3}) then implies
that $\dot{\mu}$ is
also a function of $\mu$.

Now, the right hand-side of the Gibb's relation,
$\frac{1}{T} d\left(\frac{\mu}{n} \right) \,+\, \frac{p_t}{T}
d\left(\frac{1}{n}  \right)$ (where $n$ is
the baryon number density), is a perfect differential, which implies that
either $n = n(\mu)$ (whence, from the baryon conservation law,
$\dot{n} = - n \theta\,,\, \dot{n}$ and hence $\dot{\mu}$ is a function of
$\mu$ only), or we
have the integrability condition
\beq
    T\, \frac{dp_t}{d\mu} \;=\; ( \mu \,+\, p_t)\, \frac{dT}{d\mu}
       \;\;\;, \label{8.6}
\eeq
whence from (\ref{8.5}) we obtain
\beq
    \theta \;=\; -\frac{\mu \,+\, p_t}{\chi \,T} \,
		 \left(\frac{dp_t}{d\mu}\right)^{-1} \;\;\;. \label{8.7}
\eeq
Assuming that $\frac{dp_t}{d\mu} > 0$ (and noting that $(\mu \,+\, p_t)
(\chi T)^{-1} >0$ if the weak energy condition is
satisfied) we thus have $\theta <0$.  Otherwise (\ref{8.7}) can be
regarded as an expression defining $\chi$.

Summarising, we have that
\bea
    \theta &=& f_1(\mu) \;\;\;, \label{8.8a} \\
    \dot{\mu} &=&  f_2(\mu) \;\;\;, \label{8.8b}
\eea
[in addition to $p = f_3(\mu)$].  These imply the conditions
\beq
    \frac{\p f_i}{\p x^\alpha}\, \frac{\p \mu}{\p t} \;=\;
    \frac{\p f_i}{\p t}\, \frac{\p \mu}{\p x^\alpha} \;\;\;;
        \qquad i = 1, 2(3) \label{8.9}
\eeq
which imply conditions on the (arbitrary functions in the) metric
(\ref{3.11})--(\ref{3.12})
using (\ref{3.15}), (\ref{3.16}) and (\ref{3.17}).

For example, in the case of spherical symmetry, when the functions
$b_\alpha(t) =0$ in (\ref{3.12}) whence the metric is given by (\ref{1.2}),
and $\mu, p$ and $q$,
are given by equations (\ref{1.4}), (\ref{1.5}) and (\ref{1.6}),
respectively, (\ref{8.8a}) implies
that
\beq
    F(t) \,  G(t) \;=\; \alpha \;\;\;, \label{8.10}
\eeq
where $\alpha$ is a constant, whence
\beq
    \mu \;=\;
	   \frac{3(F^2 \, r^2 \,-\, \alpha)^2}{(F^2 \, r^2 \,+\, \alpha)^2}
	   \, \frac{\dot{F}^2}{F^2} \,+\, 12\, \alpha \;\;\;.
\eeq
Equations (\ref{8.8b}) and (\ref{8.9})
$(\dot{\mu}_{, r}\, \mu_{,t} = \dot{\mu}_{,t}\, \mu_{,r})$ then imply that
\beq
    \ddot{F} F \;=\; \dot{F}^2 \;\;\;,
\eeq
and hence
\beq
    F \;=\; c\, e^{\beta t} \;\;\;, \label{8.11}
\eeq
where $\beta$ and $c$ integration constants.
A constant rescaling of $r$ can then
be used to set $c = \alpha$, whence the metric can be written as
\beq
     ds^2 \;=\; -dt^2 \,+\, \frac{dr^2 \,+\, r^2 d\theta^2 \,+\, r^2
                \sin^2\theta \, d\phi^2}
               {[\alpha \, r^2 \, e^{\beta t} \,+\, e^{-\beta t}]^2}
		 \;\;\;. \label{8.12}
\eeq
Spacetimes of this form have been studied before \cite{COLEY&TUPPER_B}.
{}From equations (\ref{1.4})--(\ref{1.6}) we can then write
\bea
    \mu &=& 12\, \alpha \;\,+\,\; \frac{1}{3} \, \theta^2 \;\;\;,\label{8.13a}
\\
      p &=&  2\,  \beta^2 \,-\, 4\, \alpha \,-\, \frac{5}{9}\, \theta^2
      \;\;\;, \label{8.13b}
\eea
where
\beq
    \theta \;=\; 3\, \beta\, (1 \,-\, \alpha\, e^{2 \beta t}\, r^2)
		 (1 \,+\, \alpha\, e^{2  \beta t}\, r^2)^{-1}
		 \;\;\;, \label{8.13c}
\eeq
and
\beq
    q_1 \;=\; - \frac{8\, \alpha\, \beta\, r}{[\alpha \, r^2 \, e^{\beta t}
                      \,+\, e^{-\beta t}]^2}  \;\;\;. \label{8.13d}
\eeq

%%%%%%%%%%%%%%%%%%%%%%%%%%%%%%%%%%%%%%%%%%%%%%%%%%%%%%%%%%%%%%%%%%%%%%%%
%\newpage

\noindent{\bf Perfect Fluid plus a Magnetic Field}

\noindent
If the source of the gravitational field is a combination of a perfect
fluid and a pure magnetic field then the stress-energy tensor
has the form
\beq
    T_{ab} \;=\; \overline{\mu}\, u_a\, u_b \,+\, \overline{p}\, (g_{ab}
                 \,+\, u_a \, u_b) \,+\, T^{MAG}_{ab} \;\;\;. \label{8.14}
\eeq
If $h^a$ is the local magnetic field measured by $u^a$ (and the
local electric field $e^a$ is zero) then the stress-energy tensor for a pure
magnetic field is given by
the Minkowski tensor \cite{LICHNEROWICZ}
\beq
    T^{MAG}_{ab} \;=\; \lambda\, \left[ \left(\frac{1}{2}\, g_{ab} \,-\,
                       u_a \, u_b \right) \, h^2 \,-\, h_a \,h_b \right]
                       \;\;\;, \label{8.15}
\eeq
where $\lambda$ is the magnetic permeability (assumed constant) and $h^2 \equiv
h_a h^a$ ($h^a$ is orthogonal to
$u^a$, $h^a u_a = 0$).  For instance, the above stress-energy tensor,
(\ref{8.14}), would be appropriate for a plasma in a strong magnetic field when
the particle collision density is low \cite{CHEW,TENERIO&HAKIM}.

Equation (\ref{8.14}) is formally of the form of an anisotropic fluid
\cite{MAARTENS&MASON}  with energy density and isotropic pressure
\bea
    \mu &=& \overline{\mu} \,+\, \frac{\lambda}{2} \, h^2 \;\;\;,
\label{8.16}\\
      p &=& \overline{p} \,+\, \frac{\lambda}{6}\, h^2 \;\;\;, \label{8.17}
\eea
and anisotropic stress tensor
\beq
    \pi_{ab} \;=\; \lambda\, h^2 \,\left\{ \frac{1}{3} \,(g_{ab} \,+\, u_a
                   \,u_b) \,-\, n_a\,  n_b \right\}\;\;\;, \label{8.18}
\eeq
where $n_a = h_a/h$.  The anisotropic pressures are formally
\bea
    p_\perp &=& \overline{p} \,+\, \frac{\lambda}{2}\, h^2\;\;\;,
\label{8.19}\\
    p_\parallel & =& \overline{p} \,-\, \frac{\lambda}{2}\, h^2
                     \;\;\;. \label{8.20}
\eea

The magnetic field must also satisfy Maxwell's equations which
reduce to
\beq
    (u^i\, h^j \,-\, u^j\, h^i)_{;i} \;=\; 0 \label{8.21}
\eeq
for a pure magnetic field \cite{LICHNEROWICZ}.  If the integral curves of
$u^i$ are  shear-free, irrotational and
geodesic then (\ref{8.21}) implies
\bea
    & &  h^i{}_{;i} \;=\; 0 \;\;\;, \label{8.22} \\
    & & \frac{2}{3} \, \theta\, h_i \,+\, h_{i;j}\, u^j \;=\; 0
               \;\;\;. \label{8.23}
\eea
Contracting (\ref{8.23}) with
$h^i$ implies that
\beq
    \frac{1}{2}\, \p_t (h^2) \,+\, \frac{2}{3} \, \theta \, h^2 \;=\; 0
        \;\;\;. \label{8.24}
\eeq
Now, equation (\ref{6.6}) implies that
\beq
    - \lambda^2 \, h^2 \;=\; P(x^\alpha)/H^2 (t) \;\;\;. \label{8.25}
\eeq
Inserting the result (\ref{8.25}) into (\ref{8.24}) yields $\dot{H} =0$,
and thus the expansion is zero.  Therefore, the magnetic field is static.
In comoving coordinates, (\ref{5.8}), equation (\ref{8.23}) is trivially
satisfied and (\ref{8.22}) reduces to
\beq
    \nabla^\alpha h_\alpha \;=\; 0 \;\;\;. \label{8.26}
\eeq
In addition, $h_\alpha$ must also satisfy
$\nabla^\beta \pi_{\alpha \beta} = \frac{1}{3} \nabla_\alpha  \mu$,
which reduces to
\beq
    \frac{1}{3 \lambda} \, \nabla_\alpha \overline{\mu} \;=\;
        \frac{1}{3}\, h\, \nabla_\alpha h \,-\,
         h^\beta \nabla_\beta h_\alpha \;\;\;.
\eeq
We note that there is no non-trivial solution if $\overline{p}$ is given by
the equation of state $\overline{p} = (\gamma -1) \overline \mu$ with
$\gamma$ constant and in which the weak energy condition holds.

\setcounter{equation}{0} % Reset the equation counter

%%%%%%%%%%%%%%%%%%%%%%%%%%%%%%%%%%%%%%%%%%%%%%%%%%%%%%%%%%%%%%%%%%%%%%%%%%%%%%%
%                                                                             %
                             \section{Two Component Fluids}                   %
%                                                                             %
%%%%%%%%%%%%%%%%%%%%%%%%%%%%%%%%%%%%%%%%%%%%%%%%%%%%%%%%%%%%%%%%%%%%%%%%%%%%%%%

Here, we consider the stress-energy tensor
associated with a mixture of two fluids.  We take one of the fluid components
to be a perfect fluid whose flow forms a SIG timelike congruence
(with fluid four-velocity $u^a = \delta^a_0$);
the density and pressure of this perfect fluid are denoted by $\mu_1$ and
$p_1$, respectively.  We denote
the contribution of this fluid to the stress-energy tensor by
\beq
    {}^{(1)}T_{ab} \; \equiv\; (\mu_1 \,+\, p_1)\, u_a \,u_b \,+\, p_1\, g_{ab}
      \;\;\;. \label{9.1}
\eeq
The second fluid will be taken to be either another perfect fluid with a
velocity $v^a$ that is tilted with respect to $u^a$ or a pure
radiation field with the following, respective, contributions to
the stress-energy tensor
\bea
    {}^{(2)}T_{ab} &=& (\mu_2 \,+\, p_2 ) \,v_a\, v_b \,+\, p_2 \,g_{ab}
                        \;\;\;, \label{9.2} \\
    {}^{(2)}T_{ab} &=& e\, k_a \, k_b \;\;\;. \label{9.3}
\eea
where $v^a$ is a unit timelike vector $(v^a \ne u^a)$, and
$k^a$ is a null vector, $k_a k^a =0$.  The total stress-energy tensor is simply
\beq
    T_{ab}\;=\; {}^{(1)}T_{ab} \,+\, {}^{(2)}T_{ab} \;\;\;. \label{9.4}
\eeq
The stress-energy tensor (\ref{9.4}) is equivalent to
a single fluid with a non-zero heat flux and a non-zero
anisotropic stress tensor.  However, the anisotropic stress
tensor associated with the fluid velocity $u^a$ is directly
related to the heat flux (relative to $u^a$) via an
equation of the form:
\beq
    \pi_{ab} \;=\; \pi \left\{q_a\, q_b \,-\, \frac{1}{3} \,(q_c\, q^c)\,
                   (g_{ab} \,+\, u_a \, u_b)\right\} \;\;\;. \label{9.5}
\eeq
In comoving coordinates $(u^a = \delta^a_0)$, the
non-zero components of the anisotropic stress tensor are
\beq
    \pi_{\alpha\beta} \;= \; \pi \,\left\{ q_\alpha\, q_\beta \,-\,
                           \frac{1}{3} \, (h^{\gamma\delta} \, q_\gamma
                           q_\delta) \, h_{\alpha\beta} \right\}
                           \;\;\;. \label{9.6}
\eeq

Letelier \cite{LETELIER} has examined a two-perfect-fluid model of an
anisotropic fluid.  In particular, he examined the case where the stress-energy
tensor consists of the sum of two perfect fluids, and of one perfect fluid
and a null fluid.  He studied the two-perfect-fluid model in the instance
where both the perfect fluid components were irrotational.  Ferrando,
Morales and Portilla \cite{FMP_89} have studied the two-perfect-fluid
model where one of the fluid components was shear-free,
irrotational and geodesic.  They used an initial value formulation of general
relativity to construct solutions.  We essentially rederive their
result using straightforward index notation, and indicate that the
form of the metric they derive is appropriate for all shear-free,
irrotational, geodesic timelike congruences when the anisotropic stress
tensor has the form specified by equation (\ref{9.5}).  The
results follow quite quickly from the general analysis of section 2.

Inserting the expression (\ref{9.6}) for the anisotropic stress tensor
into the left hand-side of (2.38) and then using (\ref{2.36})
to get an expression for $\p_t q_\alpha$ yields
\beq
    \nabla_\beta q_\alpha \;=\; A_\alpha\, q_\beta \,+\, A_\beta \,
                                q_\alpha \,+\, B\, q_\alpha\, q_\beta \,+\,
                                C\, h_{\alpha \beta} \label{9.7}
\eeq
(the precise expressions for $A_\alpha$, $B$ and $C$ are unnecessary
for the remainder of the discussion).  The unit spacelike vector
\beq
    w_\alpha \;\equiv\;  \frac{q_\alpha}{\sqrt{h^{\beta\gamma} \, q_\beta
                         q_\gamma}}  \label{9.8}
\eeq
then satisfies
\beq
    \nabla_\beta w_\alpha \;=\; \frac{1}{2} \,\Theta \,(h_{\alpha \beta}
              \,-\, w_\alpha\, w_\beta) \,+\, a_\alpha\, w_\beta
              \;\;\;, \label{9.9}
\eeq
where
\bea
    \Theta &=& \nabla^\alpha w_\alpha \;\;\;, \label{9.10} \\
    a_\alpha &=& w^\beta\, \nabla_\beta w_\alpha \;\;\;
                 [ = A_\alpha - (h^{\beta \gamma}\, A_\beta\, w_\gamma)\,
                 w_\alpha]\;\;\;. \label{9.11}
\eea
Therefore, the vector $w_\alpha$ is shear-free and twist-free, in terms
of the three-dimensional geometry.  Thus, the three spaces admit an
umbilical foliation \cite{JIANG}, and
hence, there exists a coordinate system such that the three-metric has the
following form \cite{FERRANDO_92,EISENHART1}
\beq
    h_{\alpha\beta}\, dx^\alpha \,dx^\beta \;=\; a^2(x^\alpha)\, dx^2 \,+\,
            b^2(x^\alpha)\, (dy^2 \,+\, dz^2) \label{9.12}
\eeq
and the vector
\beq
    w_\alpha \;=\; a\, \delta^x_\alpha \;\;\;. \label{9.13}
\eeq
Equations (\ref{9.13}) and (\ref{2.33}) imply that
$H = H(t, x)$.  Equation (\ref{2.34}) then implies that
\bea
    {}^3R_{xy} &=& \nabla_x \nabla_y ({\rm ln} H) \equiv -\p_x ({\rm ln}  H)\,
                \p_y({\rm ln} a) \;\;\;, \label{9.14} \\
    {}^3R_{xz} &=& \nabla_x \nabla_z ({\rm ln} H) \equiv -\p_x ({\rm ln}  H)\,
                \p_z({\rm ln} a) \;\;\;. \label{9.15}
\eea
But $\p_x({\rm ln} H)$ is a function of both
$t$ and $x$ (otherwise $q_\alpha$ is
identically zero).  Therefore, we must have ${}^3R_{xy} = {}^3R_{xz} = 0$ and
hence,
we can
take $a =1$ without loss of generality.   Calculating ${}^3R_{xy}$ and
${}^3R_{xz}$ for the metric (\ref{9.12}) with $a =1$, we find
\bea
    {}^3R_{xy} &=& -b^{-2}\, \p_{xy} b \,+\, b^{-3}\, \p_x b\, \p_y b
                \;\;\;, \label{9.16} \\
    {}^3R_{xz} &=&  -b^{-2}\, \p_{xz} b \,+\, b^{-3}\, \p_x b \, \p_z b
                 \;\;\;, \label{9.17}
\eea
and hence
\beq
    b \;=\; f(x)\, \phi(y,z) \label{9.18}
\eeq
for some, as yet
unknown, functions $f$ and $\phi$.  Thus, the metric for the two component
fluid (\ref{9.4}) may be written as
\beq
    ds^2 \;=\; -dt^2 \,+\, H^2(t, x)\, \{dx^2 \,+\, f^2(x) \,\phi^2(y,z)\,
                (dy^2 \,+\, dz^2)\} \;\;\;. \label{9.20}
\eeq
The only non-zero components of the anisotropic stress tensor, (\ref{2.34}),
for the above metric are
\bea
    \pi_{xx} &=& -2f^{-2} \,\phi^{-2}\, \pi_{yy} \;=\; -2f^{-2} \,\phi^{-2}\,
                  \pi_{zz} \label{9.21} \\
    &=& \frac{2}{3}\, [ -H^{-1}\, H_{xx} \,-\, 2H^{-2}\, H^2_x \,-\,
         H^{-1}\, H_x\, f^{-1}\, f_x \,-\, f^{-1}\, f_{xx} \,+\,
         f^{-2}\, f^2_x  \nonumber \\
    & &    +\,f^{-2}\, \{ \phi^{-3}\, (\phi_{yy} \,+\, \phi_{zz}) \,-\,
           \phi^{-4}\, (\phi^2_y \,+\, \phi^2_z)\}] \;\;\;. \label{9.22}
\eea
We note that $H$ is not a separable function, $H \ne X(x) T(t)$,
otherwise the energy flux and the anisotropic stress tensor would both be zero.

Finally, $\pi_{xx}$ must also satisfy
\beq
    \pi_{xx} \;=\; \frac{2}{3} \,\pi\, q^2 \label{9.23} \;\;\;,
\eeq
where
\beq
    q \;=\; 2 \p_t \p_x \,({\rm ln} H) \label{9.24}
\eeq
and the energy flux is simply $q_\alpha = q \delta^x_\alpha$.  The
functions $q$ and $\pi$ are therefore determined by (\ref{9.23}) and
(\ref{9.24}).  No further progress can be made in general, without specifying
some sort of equation of state, usually an equation relating $q$ and $\pi$.
For illustrative purposes, we will consider (i) a perfect fluid and a null
fluid
mixture, and (ii) a two-perfect-fluid mixture.

%%%%%%%%%%%%%%%%%%%%%%%%%%%%%%%%%%%%%%%%%%%%%%%%%%%%%%%%%%%%%%%%%%%%%%%%%%%

\noindent{\bf Perfect fluid plus pure radiation}

\noindent
The stress-energy tensor for a mixture of a perfect fluid and pure
radiation is
\beq
    T_{ab} \;=\; (\mu_1 \,+\, p_1)\, u_a\, u_b \,+\, p_1\, g_{ab} \,+\,
                 e\, k_a\, k_b\;\;\;, \label{9.25}
\eeq
where $k_a$ is a null vector.  We are free to rescale $k_a$ and $e$
such that $k_a u^a = -1$.  Thus, in comoving coordinates we choose
$k_a = (-1, k_\alpha)$ and $k_\alpha = H \delta^x_\alpha$.  Then,
the various quantities $(\mu, p, q_\alpha, \pi_{\alpha \beta}$) are
\bea
    \mu &=& \mu_1 \,+\, e \;\;\;, \label{9.26} \\
      p &=& p_1 \,+\, \frac{1}{3} \, e \;\;\;, \label{9.27} \\
    q_\alpha &=& e\, H\, \delta^x_\alpha \;\;\;, \label{9.28} \\
    \pi_{\alpha\beta} &=& e \,H^2\, \{\delta^x_\alpha\, \delta^x_\beta \,-\,
                          \frac{1}{3}\, h_{\alpha\beta}\} \;\;\;. \label{9.29}
\eea
Therefore,
\beq
    q  \;=\; e\, H \;=\; 2 \p_t \p_x ({\rm ln} H) \label{9.30}
\eeq
and
\beq
    \pi_{xx} \;=\; \frac{2}{3} e\, H^2 \;\;\;. \label{9.31}
\eeq
Using (\ref{9.30}), we obtain
\beq
    \pi_{xx} \;=\; \frac{4}{3} \,H \p_t \p_x ({\rm ln} H) \label{9.32}
\eeq
Comparing (\ref{9.32}) to (\ref{9.22}) we find
\beq
    \phi^{-3}\, (\phi_{yy} \,+\, \phi_{zz}) \,-\, \phi^{-4}\,
                  (\phi^2_y \,+\, \phi^2_z) \;=\; -k \label{9.33}
\eeq
for some constant $k$, and thus
\beq
    \phi^{-1} \;=\; 1 \,+\, \frac{k}{4}\, (y^2 \,+\, z^2)\;\;\;. \label{9.34}
\eeq
The functions $H$ and $f$ must also satisfy
\bea
    2H \,\p_t (H^{-1} H_x) \,+\, H^{-1}\, H_{xx} &+&
             2H^{-2}\, H^2_x \,+\, H^{-1}\, H_x \,f^{-1}\, f_x\;= \nonumber\\
      & &    -f^{-1}\, f_{xx} \,+\, f^{-2}\, f^2_x \,-\, k\, f^{-2}
              \;\;\;. \label{9.35}
\eea
The above equation follows directly from (\ref{9.22}) and (\ref{9.32}).
We seek solutions of (\ref{9.35}) where $H$ is not separable
$[H \ne X(x) T(t)]$.  It suffices to say that such solutions can be found for
non-separable $H$.  For example, suppose that
$f = 1 \,,\, k =0$ and $H=H(\tau)$ where $\tau = x + \alpha  t$
($\alpha$ constant)  then $H(\tau)$ satisfies the differential equation
\beq
     \frac{d^2 H}{d \tau^2} \;=\; \frac{2(\alpha H - 1)}{H(2 \alpha H + 1)}
      \left(\frac{dH}{d\tau} \right)^2  \;\;\;. \label{9.36}
\eeq
Equation (\ref{9.36}) can be  integrated to yield
\beq
    H^\prime \left(\;\equiv\; \frac{dH}{d\tau}\right) \;=\;
    \frac{(2 \alpha H + 1)^3}{c H^2}  \;\;\;, \label{9.37}
\eeq
where $c$ is an arbitrary constant. Thus, the line element
\beq
    ds^2 \;=\; - \,
                \left( \frac{dH}{H^\prime} \,-\, dx\right)^2 \,+\,
                 H^2\,(dx^2 \,+\, dy^2 \,+\, dz^2)\;\;\;, \label{9.37a}
\eeq
where $H^\prime$ is given by (\ref{9.37}) with $\alpha =1$,
is an example of a spacetime where the stress-energy tensor consists of
a perfect fluid (with 4-velocity $u^a = H^\prime \delta^a_H$) plus
pure radiation such that the perfect fluid's flow lines form a SIG timelike
congruence.

Once $\phi, f$ and $H$ are known then the functions $\mu_1, p_1$, and $e$ can
be calculated from
\bea
    \mu_1 &=& \mu \,-\, e \;\;\;, \label{9.38} \\
      p_1 &=& p \,-\, \frac{1}{3} \, e \;\;\;, \label{9.39} \\
        e &=& 2H^{-1}\, \p_t(H^{-1} H_x) \;\;\;, \label{9.40}
\eea
where $\mu$ and $p$ are given by (\ref{2.31}) and (\ref{2.32}).

%%%%%%%%%%%%%%%%%%%%%%%%%%%%%%%%%%%%%%%%%%%%%%%%%%%%%%%%

\newpage

\noindent{\bf Two-perfect-fluid   mixture}

\noindent
The stress-energy tensor for a two-perfect-fluid mixture is
\beq
    T_{ab} \;=\; (\mu_1 \,+\, p_1) \,u_a\, u_b \,+\, (p_1 \,+\, p_2) \,g_{ab}
                 \,+\, (\mu_2 \,+\, p_2)\, v_a \,v_b\;\;\;. \label{9.41}
\eeq
We shall assume that the two fluid four-velocities are not collinear,
$u_a \ne v_a$, else the stress-energy tensor would formally be that of a single
fluid with energy density $\mu_1 \,+\, \mu_2$ and
pressure $p_1 \,+\,p_2$ and hence the resulting spacetime would be FRW.
According to (\ref{2.4}) and (\ref{2.5}) the energy density and
the isotropic pressure of the mixture are
\bea
    \mu &=& \mu_1 \,+\, (\mu_2 \,+\, p_2)\, \cosh^2 \psi \,-\, p_2
             \;\;\;, \label{9.42} \\
      p &=& p_1 \,+\, p_2 \,+\, \frac{1}{3}\, (\mu_2 \,+\, p_2)\,
            \sinh^2 \psi \;\;\;, \label{9.43}
\eea
where
$\cosh \psi \equiv -v_a u^a$ and $\psi$ is called the tilt angle.
The energy flux, (\ref{2.6}), is
\beq
    q_a \;=\; (\mu_2 \,+\, p_2)\, \cosh \psi \,(v_a \,+\, \cosh \psi \,u_a)
               \label{9.44}
\eeq
and the anisotropic stress tensor, (\ref{2.7}), is
\beq
    \pi_{ab} \;=\; \frac{1}{\cosh^2 \psi\, (\mu_2 \,+\, p_2)}\,
                   \left[ q_a \, q_b \,-\, \frac{1}{3}\, (g^{cd}\, q_c\, q_d)
                   (g_{ab} \,+\, u_a\, u_b)\right] \;\;\;. \label{9.45}
\eeq
In comoving coordinates, $u_a = - \delta^0_a$, whence the four-velocity of the
tilting perfect fluid is given by
$v_a = (-\cosh \psi$, $\sinh \psi\, \delta^x_\alpha)$, and
\beq
    q_\alpha  \;=\; (\mu_2 \,+\, p_2)\, \cosh \psi\, \sinh \psi \,
                    \delta^x_\alpha \;\;\;, \label{9.46}
\eeq
and
\beq
    \pi_{\alpha \beta} \;=\; \frac{1}{\cosh^2 \psi (\mu_2 \,+\, p_2)}\,
                             \left[q_\alpha \, q_\beta \,-\, \frac{1}{3}\,
                             (h^{\gamma\delta}\, q_\gamma\, q_\delta)\,
                              h_{\alpha\beta}\right] \label{9.47} \;\;\;.
\eeq
Therefore, the quantities $q$ and $\pi$ are given by
\bea
    q &=& (\mu_2 \,+\, p_2)\, \cosh \psi \, \sinh \psi \;\;\;, \label{9.48} \\
    \pi &=& \frac{1}{(\mu_2 \,+\, p_2) \cosh^2 \psi} \;\;\;. \label{9.49}
\eea
{}From (\ref{9.48}) and (\ref{9.49}), we obtain the identity
\beq
    q \pi \;=\; \frac{\sinh \psi}{\cosh \psi} \;\;\;. \label{9.50}
\eeq
Now, the four quantities $\mu, p,q$, and $\pi$ are given,
respectively, by equations (\ref{2.31}), (\ref{2.32}), (\ref{9.24})
[see also (\ref{2.33})] and (\ref{9.23})
[where $\pi_{xx}$ is given by  (\ref{9.22}); see also
(\ref{2.34})] in terms of the functions $H, f$ and $\phi$ [in metric
(\ref{9.20})] and their derivatives.  In addition, $\mu, p, q$ and $\pi$ are
given in terms of the five (unknown) physical quantities
$\mu_1, p_1, \mu_2, p_2$ and
$\psi$ through equations (\ref{9.42}), (\ref{9.43}), (\ref{9.48}) and
(\ref{9.49}), respectively.  At this point, therefore, the system of (four)
equations is underdetermined, and no further progress can be made until
physical conditions on $\mu_1, p_1, \mu_2, p_2$ and $\psi$ are
specified.  If one such condition is specified, the remaining four physical
quantities can then in principle be expressed on terms of $H, f$ and
$\phi$ and their derivatives.  Two conditions
on $\mu_1, p_1, \mu_2, p_2$ and $\psi$ would then [through equations
(\ref{2.31}), (\ref{2.32}), (\ref{9.23}) and (\ref{9.24})] give rise to a
(differential) equation in terms of $H, f$ and $\phi$ that would need to be
satisfied (further restricting the form of the metric).

Let us consider the following conditions:  (i)  separate
energy conservation,  (ii)  linear equations of state for $p_1$ and
$p_2$,  (iii)  a constant tilt angle $\psi$,  (iv)  the second
perfect fluid is due to a scalar field, and  (v)  a single tilting perfect
fluid.

\noindent
(i) \, {\it Separate energy conservation}.  Here we assume that the energy
momentum of each perfect fluid is separately conserved; that is,

\noindent
$^{(A)} T_{ab}{}^{;b} = 0 \;\; (A = 1,2)$.  Due to total stress-energy
conservation, this leads to
one extra constraint:
\beq
    \frac{\dot{\mu_1}}{\mu_1 \,+\, p_1} \;=\; \frac{3 \dot{H}}{H}
                     \;\;\;, \label{9.51}
\eeq
where $p_1$ is a function of $t$ only, $p_1 = p_1(t)$.  We note
that if $\mu_1$ and $p_1$ satisfy an equation of state of the form
$\mu_1 = \mu_1(p_1) = \mu_1 (t)$, we
immediately have that $q_a = \pi_{ab} = 0$ and the
resulting spacetime is FRW.

%--------------------------------------------------------------------

\noindent
(ii) \, {\it Linear equations of state.}  Here we assume that each fluid
satisfies a linear equation of state, viz.,
\bea
    p_1 &=& (\gamma_1 \,-\, 1) \, \mu_1 \;\;\;, \label{9.52a} \\
    p_2 &=& (\gamma_2 \,-\, 1) \, \mu_2 \;\;\;, \label{9.52b}
\eea
where the $\gamma_1$ and $\gamma_2$  are constants.  Equations (\ref{9.42})
and (\ref{9.43}) then yield
\beq
    p \,+\, (1 \,-\, \gamma_1)\, \mu \;=\; \mu_2\, \{\gamma_1 \,
         (\gamma_2 \,-\, 1) \,+\, \frac{1}{3}\, \gamma_2 [-3\, \gamma_1\,
         \cosh^2 \psi \,+\, 4 \cosh^2 \psi \,-\, 1] \} \;\;\;, \label{9.53}
\eeq
and equations (\ref{9.49}) and (\ref{9.50}) imply
\beq
     \mu_2 \;=\; \frac{1}{\gamma_2 \pi \cosh^2 \psi} \qquad {\rm and} \qquad
     \frac{1}{\cosh^2 \psi} \;=\; 1 \,-\, q^2\, \pi^2 \;\;\;, \label{9.54}
\eeq
whence
\beq
    1 \,-\, \frac{\gamma_1}{\gamma_2}  \;=\; \pi\, \left\{p\,+\,
           (1 \,-\,\gamma_1) \, \mu \,-\, q^2 \, \pi \left[\frac{1}{3} \,-\,
           \gamma_1 \,+\, \frac{\gamma_1}{\gamma_2} \right] \right\}
             \;\;\;. \label{9.55}
\eeq

%--------------------------------------------------------------------

\noindent
(iii) {\it  Constant tilt angle.}  If $\psi = \psi_0 =$ constant then
from (\ref{9.50}) we have that
\beq
    q \, \pi \;=\; \frac{\sinh \psi_0}{\cosh \psi_0} \;=\; \alpha\;\;\;;
                 \qquad \alpha \quad {\rm const} \;\;\;, \label{9.56}
\eeq
whence from (\ref{9.23}) and (\ref{9.24}) we have that
\beq
    \pi_{xx} \;=\; \frac{4 \alpha}{3}\, \p_t \p_x ({\rm ln} H)
                 \;\;\;, \label{9.57}
\eeq
which is similar to the expression given by (\ref{9.32}) (with H `replaced by
$\alpha$').  In particular, $\pi_{xx}$ is a function of $t$ and
$x$ only.  Therefore, $\phi$ again satisfies the differential equation
(\ref{9.33}) for a constant $k$, and is consequently given by equation
(\ref{9.34}) [that is, $\phi^{-1} = 1 \,+\, \frac{k}{4} (y^2 \,+\, z^2)]$, and
now  $H$ and $f$ satisfy an equation identical to (\ref{9.35}) except that $H$
is `replaced by $\alpha$' in the first term on the left hand-side
[that is, the first term on the left hand-side of equation (\ref{9.35}) is now
$2 \alpha \partial_t (H^{-1} H_x)]$.

%--------------------------------------------------------------------

\noindent
(iv) \, {\it Perfect fluid plus a scalar field.}   The energy-momentum tensor
of a scalar field is formally equivalent to that of a perfect fluid
with
\bea
    \mu_\phi &=& \frac{1}{2}\, (\nabla \phi)^2 \,+\, V(\phi)
                   \;\;\;, \label{9.58a} \\
    p_\phi &=& \frac{1}{2}\, (\nabla \phi)^2 \,-\, V(\phi)
                       \;\;\;, \label{9.58b}
\eea
where $V$ is the potential of the scalar field.  Therefore, a perfect fluid
plus scalar field source can be formally treated as a two-perfect-fluid
mixture (with separate energy conservation).  The field $\phi$
satisfies a
(Klein-Gordon) differential equation; therefore, for a
specific $V(\phi)$ we will obtain a relation (effective `equation of state')
between $\mu_\phi$ and $p_\phi$.

%--------------------------------------------------------------------

\noindent
(v) \, {\it Single tilting perfect  fluid.}  Here we assume that
\beq
    \mu_1 \;=\; p_1 \;=\; 0 \;\;\;, \label{9.59}
\eeq
so that the energy-momentum tensor represents a single
perfect fluid whose four-velocity is tilting
with respect to the shear-free, irrotational and geodesic timelike
congruence (and hence the results at the end of section 3 do not apply).

{}From (\ref{9.42}) and (\ref{9.43}), using (\ref{9.59}), we have that
\beq
    \mu \,+\, p \;=\; \frac{1}{3}\, (\mu_2 \,+\, p_2)\, [3 \cosh^2 \psi \,+\,
             \sinh^2 \psi] \;\;\;, \label{9.60}
\eeq
whence from (\ref{9.48}) and (\ref{9.49}) we obtain
\bea
    q &=& \frac{3(\mu\,+\,p) \,\cosh \psi \,\sinh \psi}{3 \cosh^2 \psi \,+\,
                   \sinh^2 \psi} \;\;\;, \label{9.61} \\
    \pi &=& \frac{3 \cosh^2 \psi \,+\, \sinh^2 \psi}{3 (\mu \,+\, p) \,
                \cosh^2 \psi} \;\;\;, \label{9.62}
\eea
and hence, we have the identity
\beq
    q^2 \;=\; \pi\, q^2\, [ (\mu \,+\, p) \,-\, \frac{1}{3}\, \pi\, q^2]
                \;\;\;. \label{9.63}
\eeq
Now, from (\ref{2.31}) and (\ref{2.32}) we obtain [for metric (\ref{9.20})]
\bea
    \mu \,+\, p &=& -2H^{-1}\, H_{tt} \,+\, 2 H^{-2}\, H_t^2 \,+\,
            \frac{2}{3}\, H^{-2}\, \left\{ -2 H^{-1}\, H_{xx} \,+\,
            H^{-2} \right. \,H_x^2  \nonumber\\
     & & \left.  -4 H^{-1}\, H_x\, f^{-1}\, f_x \,-\, 2f^{-1}\, f_{xx} \,-\,
                f^{-2} \,f_x^2 \,+\, f^{-2}\, \Phi (y, z)\right\}
                       \;\;\;, \label{9.64}
\eea
where
\beq
    \Phi(y, z) \;\equiv\; - \phi^{-3}\, (\phi_{yy} \,+\, \phi_{zz})
              \,+\, \phi^{-4} \, (\phi_y^2 \,+\, \phi_z^2)
                     \;\;\;. \label{9.65}
\eeq
Using equations (\ref{9.21})--(\ref{9.24}) and (\ref{9.64}), we see
that equation (\ref{9.63}) is of the form
\beq
    a(t, x)\, \Phi^2(y, z) \,+\, b(t, x) \,\Phi(y, z) \,+\, c(t, x) \;=\; 0
                 \;\;\;, \label{9.66}
\eeq
and hence $\Phi(y,z) = k$, a constant; consequently
$\phi$ again satisfies the differential equation (\ref{9.33})
and is thus given by equation (\ref{9.34}) [that is,
$\phi^{-1} = 1 \,+\, {k \over 4} (y^2 \,+\, z^2)].$

Finally, using (\ref{9.23}) and (\ref{9.24}), (\ref{9.63}) becomes
\beq
    \left[ \frac{H_{xt}}{H} \,-\, \frac{H_x H_t}{H^2}\right]^2 \;=\;
            \frac{3}{8} \, \pi_{xx} \, \left[ (\mu \,+\, p) \,-\, \frac{1}{2}\,
            \pi_{xx}\right] \;\;\;, \label{9.67}
\eeq
where $(\mu \,+\,p)$ is given by (\ref{9.64}), $\pi_{xx}$ is given by
(\ref{9.21}) and (\ref{9.22}), and where $\Phi(y,z) = k$.
Equation (\ref{9.67}) is a differential equation for $H(t,x)$ and $f(x)$.

 Similar to the case of a perfect fluid with pure radiation, it is possible to
find special solutions to the differential equation (\ref{9.67}). If $f=1, k=0$
and $H = H(\tau)$, where $\tau = x + \alpha t$  ($\alpha$  constant), then
(\ref{9.67}) becomes a quadratic equation in $H^{\prime\prime}/(H^\prime)^2$
($H^\prime \equiv \frac{dH}{d\tau}$):
\bea
     H^2\, \left\{(1 \,+\, 6 \alpha^2) \,H^2
           \,-\,4\right\}\left[\frac{H^{\prime\prime}}{(H^\prime)^2}
           \right]^2 &+&
           2\,H\, \left\{(2 \,-\, 15\,\alpha^{2})\,H^{2} \,-\, 3\right\}\,
           \left[\frac{H^{\prime\prime}}{(H^\prime)^2}\right]
                        \nonumber \\
    &+&4\,\left\{(1\,+\,6\, \alpha^2)\,H^2 \,+\,1 \right\} \;=\; 0
               \;\;\;. \label{9.68}
\eea
This quadratic equation can be solved and the solution for $H(\tau)$ can
be formally obtained. As an illustration,
for the specific choice of $\alpha^2 = \frac{4}{3}$,
\beq
     \frac{H^{\prime\prime}}{(H^{\prime})^2}  \;=\;
     \frac{18\,H^2+3-\sqrt {216\,H^2+25}}{H\left (9\,H^2-4\right )}
     \label{9.69}
\eeq
is one of the solutions of the quadratic equation (\ref{9.68}).
Equation (\ref{9.69}) can be integrated to yield
\beq
    H^\prime \;=\; c\,
         \frac{|9 \,H^2 \,-\, 4|^{\frac{11}{4}}\; |5\,\sqrt{216 \,H^2 \,+\, 25}
               \,-\, 25 \,-\, 108\, H^2|^{\frac{5}{8}}}
              {H^2\; |11\,\sqrt{216 \,H^2 \,+\, 25} \,-\,73\,-\,
               108\,H^2|^{\frac{11}{8}}} \;\;\;, \label{9.70}
\eeq
where $c$ is an arbitrary constant. Thus, the line element
\beq
    ds^2 \;=\; -\frac{3}{4}\left(\frac{dH}{H^\prime} \,-\, dx\right)^2
                \,+\, H^2\,(dx^2 \,+\, dy^2 \,+\,dz^2) \;\;\;, \label{9.71}
\eeq
where $H^\prime$ is given by  (\ref{9.70}),
is an example of  a model with a
single tilting perfect fluid where the fluid is
tilting with respect to a shear-free, irrotational, and
geodesic timelike congruence.

Once $\phi\,, f$ and $H$ are known, the scalar $\mu_2$ and $p_2$ can be
calculated from
\bea
    \mu_2 &=& \mu \,-\, \frac{3}{2}\, \pi_{xx} \;\;\;, \\
    p_2   &=& \frac{2 q^2}{3 \pi_{xx}} \,-\, \mu \;\;\;,
\eea
where $\mu$, $\pi_{xx}$ and $q$ are given by equations (\ref{2.31}),
(\ref{9.22}) and (\ref{9.24}), respectively.

%%%%%%%%%%%%%%%%%%%%%%%%%%%%%%%%%%%%%%%%%%%%%%%%%%%%%%%%%%%%%%%%%%%%%%%%%%%%%%%
%                                                                             %
                             \section{Discussion}                             %
%                                                                             %
%%%%%%%%%%%%%%%%%%%%%%%%%%%%%%%%%%%%%%%%%%%%%%%%%%%%%%%%%%%%%%%%%%%%%%%%%%%%%%%

In this paper, we have considered spacetimes,
with a general energy-momentum tensor [formally decomposed with respect
to $u^a$ according to (\ref{2.2})--(\ref{2.7})],
in which the
shear, vorticity and acceleration of a timelike congruence $u^a$ are all zero.
In such spacetimes, it is always possible to choose a synchronous coordinate
system with $u^a = \delta^a_t$ [see (\ref{2.19})--(\ref{2.20})] such that all
dynamical information is encoded in a single scalar function $H(t,x^\alpha)$
and such that the geometry of each of the spacelike hypersurfaces $t =$ const
is encoded (up to a conformal factor)
in the positive-definite three-metric $h_{\alpha\beta}(x^\gamma)$.

In section 3, spacetimes with zero anisotropic
stress (with respect to $u^a$) were considered.
Each three-space (of constant $t$) was shown to be
a space of constant curvature. However, the curvature was not necessarily
the same on all the $t =$ const hypersurfaces.
In section 4, spacetimes with zero heat flux vector (with respect to $u^a$)
were considered.  The scalar function $H$ was demonstrated to be independent
of the spatial coordinates [that is, $H = H(t)$], and the
anisotropic stress tensor was shown to be given entirely in terms
of $h_{\alpha \beta} (x^\gamma)$
through (\ref{5.1}).
When the energy
density and isotropic pressure satisfy a barotropic
equation of state, $p =p(\mu)$, we saw that
(except in the special case $\frac{dp}{d\mu} = -\frac{1}{3}$) the
Ricci scalar of the three-surface [with metric $h_{\alpha\beta}(x^\gamma)]$
is constant, whence the anisotropic stress-tensor is `divergence-free'
(all of the Bianchi and Kantowski-Sachs models admit subcases which fall
into this category).
In section 5,
an anisotropic fluid source,
(\ref{6.1}), was studied. These  models were shown to be a subset
of the models discussed in section 4. The fluid's
anisotropic stress tensor
was  found to possess two distinct
eigenvalues (and hence the spacetime is of Petrov type D).

In section 6, two physically relevant energy-momentum tensors
were considered, which
were subcases of models dealt with in earlier sections.  First, a viscous fluid
with heat conduction satisfying the phenomenological equations
(\ref{8.1})--(\ref{8.3}) was considered, a subcase of the models with zero
anisotropic stress tensor.  In the case
that the temperature depends only on the energy density, $T = T(\mu)$, the
various governing equations and thermodynamical laws were found to imply a
number of constraints; in particular,
we were able to `integrate' these conditions
for the  spherical symmetric case, (\ref{8.12})--(\ref{8.13d}).
Second, a perfect fluid plus a magnetic field with energy-momentum tensor given
by (\ref{8.14}) and (\ref{8.15}),
which is formally equivalent to  an
anisotropic fluid, was considered.
For this case, the Maxwell equations were shown to imply that
the expansion must be zero and hence, the magnetic field must be static.

Finally, in section 7, a number of physically relevant cases were
investigated which can be considered (or grouped together) as two
component fluids in which the stress-energy tensor is given by
(\ref{9.4}) with (\ref{9.1}) (representing a  perfect fluid whose flow forms
a SIG timelike congruence) and one of (\ref{9.2}) or
(\ref{9.3}) [representing a second (tilting) fluid].  The stress-energy tensor
is formally equivalent to that of
 a single fluid with a non-zero heat flux and a
non-zero anisotropic stress tensor that is directly related to the heat flux
via (\ref{9.5}).   In this particular case,  coordinates can be
chosen such that the metric has the simplified form
(\ref{9.20}) and only depends on three
arbitrary functions, $H(t,x)$, $f(x)$ and
$\phi(y,z)$.  A number of special cases were then considered.
First, the `second fluid' was taken to be pure radiation,
(\ref{9.3}), so that the total stress-energy tensor is
due to a mixture of a perfect fluid and pure radiation.
In this case, the function $\phi(y,z)$ is given by equation
(\ref{9.34}), and the functions $H(t,x)$ and $f(x)$ must satisfy the
differential equation (\ref{9.35}).
In the second case, the source was taken to be two non-collinear perfect
fluids,
(\ref{9.41}).
The various quantities in the formal expression for the stress-energy
tensor were then found to be given by equations (\ref{9.42}),
(\ref{9.43}), (\ref{9.48}) and (\ref{9.49}).  These four equations relate five
physical quantities to the metric functions $H, f$ and $\phi$, and their
derivatives.  Consequently, no further progress could be made until additional
conditions were imposed on the five physical quantities.
A number of conditions were then considered.
% (i)   separate conservation equations for the two perfect fluids;
% (ii)  linear equations of state  for both the perfect fluids;
% (iii) constant tilt angle  between the two fluids;
% (iv)  scalar field source for the second perfect fluid;
The case of a single perfect fluid tilting with respect to the SIG
timelike congruence
is of special interest. In particular,  we note that a
general relativistic spacetime that admits a SIG timelike congruence
and whose stress-energy tensor is  a (single)
perfect fluid need not be FRW. The metric given by
(\ref{9.71}) represents a specific example of such a spacetime.
However, the spacetime is an FRW model when the perfect fluid's flow
lines form a SIG timelike congruence.

\setcounter{equation}{0} % Reset the equation counter

\noindent{\bf Acknowledgements:}
The authors thank Robert van den Hoogen for comments on the manuscript.
This work was supported, in part, by the Natural Sciences and Engineering
Research Council of Canada.

 \end{document}